\documentclass[twocolumn,notitlepage,superscriptaddress,nofootinbib,aps]{revtex4-2}
\usepackage{amsmath,amssymb}
\usepackage{tempora}
\usepackage{newtxmath}
\usepackage[english]{babel}
\usepackage[x11names]{xcolor}
\usepackage{latexsym,amsmath,amssymb,amsbsy,graphicx,float}
\usepackage[unicode,colorlinks,citecolor=Blue3,linkcolor=Blue3,bookmarks=true]{hyperref}

\newcommand{\corr}[1]{#1}
\addto\captionsrussian{}

\begin{document}

\title{Dynamics of squeezed-state preparation in quadratic and Kerr nonlinear microresonators}

\author{S.D. Chikalkin}
\email{sergey.chikalkin@gmail.com}
\affiliation{Russian Quantum Center, Skolkovo 121205, Russia}
\affiliation{Moscow Institute of Physics and Technology, 141701 Dolgoprudny, Russia}
\author{V.L. Gorshenin}
\affiliation{Russian Quantum Center, Skolkovo 121205, Russia}
\affiliation{Moscow Institute of Physics and Technology, 141701 Dolgoprudny, Russia}
\author{V.E. Lobanov}
\affiliation{Russian Quantum Center, Skolkovo 121205, Russia}
\author{I.A. Bilenko}
\affiliation{Russian Quantum Center, Skolkovo 121205, Russia}
\affiliation{Faculty of Physics, Lomonosov Moscow State University, 119991 Moscow, Russia}
\author{D.A. Chermoshentsev}
\email{d.chermoshentsev@gmail.com}
\affiliation{Russian Quantum Center, Skolkovo 121205, Russia}
\affiliation{Moscow Institute of Physics and Technology, 141701 Dolgoprudny, Russia}

\begin{abstract}
\noindent We develop a dynamical model of squeezed-state generation in nonlinear optical microresonators that takes into account the back-action of the signal mode on the pump mode. The resonator is treated as an open quantum system, while the pump modes are described by time-dependent classical amplitudes, which allows pump depletion to be included beyond the standard undepleted-pump approximation. For a $\chi^{(2)}$ microresonator, the model reproduces the conventional 3~dB intracavity squeezing limit below threshold and predicts a transient above-threshold regime in which this limit can be overcome for a finite time before pump depletion becomes appreciable. The dynamical approach is extended to a three-mode Kerr microresonator with two pump modes. Separately, we investigate stationary squeezing in a reduced five-mode model with fixed pump amplitudes, including a parasitic single-pump spontaneous four-wave-mixing channel. Within this model, the parasitic channel reduces the achievable squeezing and shifts the instability threshold to lower pump amplitudes. These results provide a compact framework for estimating the validity range of the undepleted-pump approximation and for analyzing transient squeezing dynamics in nonlinear microresonators.
\end{abstract}

\maketitle

\section{Introduction}

The generation and usage of squeezed states of light - quantum states with reduced noise in one field quadrature below the vacuum level - play a key role in modern quantum optics and precision measurement~\cite{Schnabel2017}. Such states enable sensitivity enhancement in quantum metrology \cite{Giovannetti2011}, gravitational-wave interferometry \cite{Caves_PRD_23_8_1981}, quantum cryptography~\cite{Hillery_PRA_61_2_022309_2000}, generation of non-Gaussian states of light, such as \corr{Schr\"odinger-cat states} \cite{Dakna1997, Podoshvedov_SciRep_13_1_2023, Kuts_PhysScr_97_11_2022}, and serve as an essential resource for continuous-variable quantum information processing \cite{Rad_Nature_638_8052_912_919_2025}.  
Their practical realization therefore remains an active and highly relevant research area.

The main physical mechanism used to generate squeezed light is nonlinear optical interaction. Two common approaches are degenerate parametric down-conversion in $\chi^{(2)}$ media~\cite{Raiford_PRA_2_4_1970} and four-wave mixing via the Kerr interaction in $\chi^{(3)}$ media~\cite{Yuen_OptLett_4_10_1979}. Optical losses introduce a threshold for parametric oscillation.
Below threshold, the intracavity field can exhibit squeezing, while above threshold a degenerate parametric oscillator can also exhibit phase bistability~\cite{Okawachi_OptLett_40_22_2015,Tatarinova_PhysRevApplied_24_024010_2025}. This phase bistability has found applications in coherent optical computing~\cite{Inagaki2016} and quantum random-number generation~\cite{Marandi2012,Okawachi2016}.  

One of the most efficient ways to enhance nonlinear optical effects and generate squeezed light at lower pump powers is to use high-quality-factor (high-Q) optical microresonators~\cite{Strekalov_JOpt_18_12_123002_2016,Savchenkov_IPNPR_42_162_2004}. With the rapid development of integrated photonics, ring microresonators with ultra-high quality factors ($Q>10^6$) and strong field confinement have become widely available \cite{Gaeta_NatPhotonics_13_3_158_169_2019, Kovach_AdvOptPhotonics_12_1_135_222_2020}.  
Silicon \cite{Perez_NatComm_14_1_2023} and silicon nitride \cite{Li_Nat_639_8056_2025, Xuan_Opt_3_11_2016, Okawachi_OptLett_40_22_2015} are the most common material platforms due to their technological maturity and compatibility with CMOS fabrication.  
As a result, degenerate four-wave mixing in $\chi^{(3)}$ media has become an established approach for on-chip generation of squeezed light~\cite{Danilin_PhysRevApplied_25_054021_2026}.

Early theoretical works on degenerate parametric oscillation inside nonlinear optical cavities predicted the 3~dB squeezing limit for the intracavity field \cite{Milburn_OptComm_39_6_1981}, whereas the radiation outcoupled from the cavity can exhibit much higher degrees of squeezing \cite{Yurke_PRA_29_1_1984}.  
This external squeezing is directly measurable and provides a key resource for integrated continuous-variable photonic circuits~\cite{Masada_NatPhotonics_9_5_316_319_2015}. 
Meanwhile, the internal resonator state remains of great interest for quantum non-demolition (QND) measurements \cite{Ma_PRA_92_2_2015} and optomechanical rigidity enhancement \cite{Korobko_PhysLettA_382_33_2018}.

Despite significant experimental progress, the development of a theoretical description of squeezed-light dynamics in microresonators remains challenging. In the lossless case, exact quantum simulations of the Hamiltonian dynamics with fully quantum and depleting pump modes reveal the formation of bistable and non-Gaussian states~\cite{Nikitin_QOptB_3_2_1991, Gorshenin_JOSAB_42_2_2025}. However, including optical losses in the quantum regime makes full simulations computationally difficult.
\corr{Moreover, high-$Q$ microresonators} support a dense set of modes with close frequencies, and the excitation of additional sideband modes quickly increases model dimensionality and computational cost \cite{Chiasera_LaserPhotRev_4_3_2010}.  
Previous theoretical studies have investigated squeezing spectra~\cite{Gardiner_OptComm_50_3_1984,Collett_PRA_32_5_1985, Vaglica_OptComm_7_2_3_1990}, nonlinear state generation and transient dynamics~\cite{Yurke_PRA_32_1_1985,Krippner_PRA_50_5_1994,Hughes_PhysRevA_110_063718_2024}, and the effects of finite pump bandwidth on intracavity squeezing~\cite{Drummond_PRA_37_5_1988}. Despite these advances, the role of pump depletion in the transient dynamics of intracavity squeezing has received comparatively less attention, with pump amplitudes often treated as fixed. This assumption significantly simplifies the analysis but fails near the threshold, where pump depletion and signal-pump coupling become non-negligible.

To address these limitations, in this work we employ and extend the formalism proposed in \cite{Averchenko_OptAndSpectroscopy_105_5_2008}. This approach allows one to describe the temporal evolution of the squeezed states inside a high-Q microresonator represented as an open quantum system and to account for the back-action of the signal mode on the pump.  
By introducing a dynamic (time-dependent) classical pump amplitude rather than a constant one, the model naturally includes pump depletion and provides a consistent way to explore both below-threshold and above-threshold regimes.  
In particular, the model allows us to estimate how pump depletion limits the duration of transient intracavity squeezing beyond 3 dB.

The paper is organized as follows.  
Section~\ref{sec:quadratic_nonlinear} analyzes the formation of the squeezed state in an optical microresonator with $\chi^{(2)}$ nonlinearity.  
In Subsection~\ref{sec:under-threshold-mode}, we reproduce the well-known 3~dB squeezing limit for below-threshold operation.  
Subsection~\ref{sec:above-threshold-mode} explores the above-threshold regime and demonstrates the possibility of transient squeezing beyond 3~dB.  
Section~\ref{sec:qubic_nonlinear} extends the analysis to $\chi^{(3)}$ media, considering both idealized three-mode and more realistic multi-mode configurations.  
Finally, Section~\ref{sec:discussion} summarizes the results and outlines potential extensions of the proposed approach.

\section{Generation of squeezed states via parametric interaction}\label{sec:quadratic_nonlinear}

We first consider squeezed-state generation in a microresonator with a quadratic nonlinear medium. In the configuration studied below, a pump mode at the second harmonic and a signal mode at the first harmonic interact through spontaneous degenerate parametric interaction within the resonator. The model takes into account optical losses for the signal and pump modes. Light is also injected into the resonator at the pump mode frequency.

For simplicity, several effects are not included in the present model. In particular, we neglect material dispersion in the resonator (assuming an equidistant mode spectrum) and any detuning between the frequencies of the input pump and signal fields and the corresponding cavity resonance frequencies. However, the method we used to calculate the evolution of the light mode operators can be extended to account for these effects in future studies. 

We extend the approach of \cite{Averchenko_OptAndSpectroscopy_105_5_2008}, which allows us to consider the temporal dynamics of processes inside the resonator taking into account the inverse influence (back-action) of the signal mode on the pump mode. Our approach is based on the Lindblad equation for the system of light modes in the resonator. We assume that the pump mode remains in a coherent state. This makes it possible to obtain the time evolution of the variances of the field quadratures and the degree of squeezing of the signal mode.

The density matrix \(\hat{\rho}_T\) of the total system obeys the Lindblad equation~\cite{Gardiner_2004_Handbook_stochastic}
\begin{equation}
    \frac{d\hat{\rho}_T}{dt} = -i[\hat{H}_T, \hat{\rho}_T] + \hat{L}(\hat{a}) + \hat{L}(\hat{a}_p)\,,
    \label{eq:lindblad2total}
\end{equation}
where the Hamiltonian $\hat{H}_T$ and the dissipator $\hat{L}(\hat{a})$ are given by
\begin{equation}
    \hat{H}_T = \frac{g}{2}\Big((\hat{a}^{\dagger})^2\,\hat{a}_p + \hat{a}^2\,\hat{a}_p^{\dagger}\Big) - \varepsilon\,(\hat{a}_p^{\dagger} + \hat{a}_p)\,,
    \label{eq:hamiltonian2total}
\end{equation}
\begin{equation}
    \hat{L}(\hat{a}) = \frac{\gamma}{2}\Big([\hat{a}\hat{\rho}_T, \hat{a}^{\dagger}] + [\hat{a}, \hat{\rho}_T \hat{a}^{\dagger}]\Big)\,,
    \label{eq:collaps_op}
\end{equation}  
where $\gamma$ is the photon decay rate in the resonator, $g$ is the quadratic nonlinearity coefficient (see Appendix
\,\ref{app: chi2-coeff}), and $\hat{a}$ and $\hat{a}_p$ are the annihilation operators of the signal and pump modes, respectively. Here and throughout this work, the Hamiltonians are expressed in angular-frequency units, i.e., the physical Hamiltonians divided by $\hbar$. The external drive amplitude $\varepsilon$ for the second-harmonic pump is given by
\begin{equation}
    \varepsilon = \sqrt{\frac{P\,\gamma_c}{\hbar \omega_p}}\,,
    \label{eq:eps_expression}
\end{equation}
where $\omega_p$ is the pump frequency, $P$ is the input power in the waveguide, and $\gamma_c = \omega_p/Q_c$ is the photon decay rate due to coupling losses ($Q_c$ is the external coupling quality factor).

At the initial stage of evolution, before pump depletion becomes appreciable, we assume that the signal and pump modes can be approximately described by squeezed and coherent states, respectively. Motivated by this assumption, we represent the pump mode by a time-dependent classical amplitude, neglecting its quantum fluctuations and correlations with the signal mode. We therefore make the substitution
\begin{equation} 
    \hat{a}_p \rightarrow i\,\alpha_p(t)\,, 
    \label{eq:coherent_pump}
\end{equation}
where the averaged pump amplitude is introduced as

\begin{equation}
    \alpha_p = -\,i\,\langle \hat{a}_p(t)\rangle.
\end{equation}

\noindent
Without loss of generality, $\alpha_p$ can be taken to be real (since the phase of $\alpha_p$ is arbitrary and can be obtained by a rotation of the phase-space axes). For a coherent pump, \corr{$\alpha_p$ is the mean pump quadrature in the chosen phase convention}, where the angle brackets indicate ensemble averaging. Here the dimensionless quadrature operators are defined in the standard way:
\begin{equation}
    \hat{X} = \corr{\frac{1}{2}}(\hat{a} + \hat{a}^{\dagger})\,,
    \qquad 
    \hat{P} = \corr{\frac{1}{2i}}(\hat{a} - \hat{a}^{\dagger})\,.
    \label{eq:quad_def}
\end{equation}
\noindent
The reduced equation for the signal-mode density matrix is
\begin{equation}
    \frac{d\hat{\rho}}{dt} = -i[\hat{H}, \hat{\rho}] + L(\hat{a})\,,
\end{equation}
with an effective Hamiltonian 
\begin{equation}
    \hat{H} = i\,\frac{g\,\alpha_p}{2}\Big(\hat{a}^{\dagger 2} - \hat{a}^2\Big)\,.
    \label{eq:hamiltonian_signal}
\end{equation}
\noindent
In the Heisenberg picture, the Lindblad master equation yields the following equation of motion for an arbitrary system operator $\hat A$ \cite{Gardiner_2004_Handbook_stochastic}: 
\begin{equation}
    \frac{d\hat{A}}{dt} = i[\hat{H}, \hat{A}] + \frac{\gamma}{2}\Big(\hat{a}^{\dagger}\, [\hat{A}, \hat{a}] + [\hat{a}^{\dagger}, \hat{A}]\, \hat{a}\Big)\,.
    \label{eq:heis-lang-def}
\end{equation}
Applying Eq.~\eqref{eq:heis-lang-def} to the signal-mode quadratures and the full Lindblad equation~\eqref{eq:lindblad2total} to the pump operator $\hat{a}_p$, we obtain a closed system of equations for the quadrature variances and the pump amplitude:
\begin{equation}
    \frac{d\langle X^2\rangle}{dt} = \big( 2g\,\alpha_p - \gamma \big)\langle X^2\rangle + \frac{\gamma}{4}\,,
    \label{eq:dynamic_X2_dim}
\end{equation}
\begin{equation}
    \frac{d\langle P^2\rangle}{dt} = -\big( 2g\,\alpha_p + \gamma \big)\langle P^2\rangle + \frac{\gamma}{4}\,,
    \label{eq:dynamic_P2_dim}
\end{equation}
\begin{equation}
    \frac{d\alpha_p}{dt} = \varepsilon - \frac{g}{2}\,\big(\langle X^2\rangle - \langle P^2\rangle\big) - \frac{\gamma}{2}\,\alpha_p\,.
    \label{eq:dynamic_pump_dim}
\end{equation}
\noindent
This system can be rewritten as
	\begin{equation}
		\frac{d\langle X^2\rangle}{d\tau} 
		= 
		(\eta - 1)\langle X^2\rangle + \frac{1}{4},
		\label{eq:dynamic_X2}
	\end{equation}
	\begin{equation}
		\frac{d\langle P^2\rangle}{d\tau} 
		= 
		-(\eta + 1)\langle P^2\rangle + \frac{1}{4},
		\label{eq:dynamic_Y2}
	\end{equation}
	\begin{equation}
		\frac{d\eta}{d\tau} 
		= 
		\frac12\eta^{(0)} 
		- 
		\lambda(\langle X^2\rangle - \langle P^2\rangle) - \frac12\eta,
		\label{eq:dynamic_pump}
	\end{equation}
	where we have introduced the dimensionless variables and parameters
	\begin{equation}
		\tau = \gamma t,\,\eta = \frac{2g\alpha_p}{\gamma},\, \lambda = \frac{g^2}{\gamma^2},
	\end{equation} 
	\begin{equation}
		\eta^{(0)} = \corr{\frac{4g}{\gamma^2}\varepsilon} = \frac{2g}{\gamma}\alpha_p^{(0)}\,, \quad \corr{\alpha_p^{(0)}=\frac{2\varepsilon}{\gamma}}.
	\end{equation}
    Here $\tau$ is the time normalized to the average lifetime of a photon in the resonator, $\eta$ is the pump amplitude normalized to a threshold value,  $\lambda$ is the nonlinearity coefficient normalized to losses in the resonator, and $\eta^{(0)}$ denotes the normalized steady-state pump amplitude obtained when the nonlinear back-action of the signal mode on the pump is neglected.


  The undepleted-pump approximation applies when the back-action of the signal mode on the pump is negligible,
  
	\begin{equation}
		\lambda(\langle {X^2}\rangle -  \langle {P^2}\rangle)\ll 1.
		\label{eq:undepleting-pump-approx}
	\end{equation}
	 Neglecting the back-action term in Eq.~\eqref{eq:dynamic_pump}, we obtain the following stationary solution of Eqs.~\eqref{eq:dynamic_X2}--\eqref{eq:dynamic_pump}:
	
	\begin{equation}
		\langle {X^2}\rangle = \frac{0.25}{1-\eta^{(0)}},
		\label{eq:zero_solution_X2}
	\end{equation}
	\begin{equation}
		\langle {P^2}\rangle = \frac{0.25}{\eta^{(0)} + 1},
		\label{eq:zero_solution_Y2}
	\end{equation}
	\begin{equation}
		\eta = \eta^{(0)}.
		\label{eq:zero_solution_pump}
	\end{equation}

    \noindent
	  Threshold value \(\eta^{(0)} = 1\) physically corresponds to sufficient pumping mode power to generate phase bistable states in the signal mode \cite{Averchenko_OptAndSpectroscopy_105_5_2008}. Within the undepleted-pump approximation, a stable stationary solution exists only below the threshold, $\eta^{(0)}<1$. Above threshold, $\eta^{(0)}>1$, the variance $\langle X^2\rangle$ grows indefinitely within this approximation, indicating the need to account for pump depletion.

    The coefficient of nonlinearity \(g\) for the degenerate parametric interaction and the Q-factor can be estimated theoretically (see Eq.\,\eqref{eq:non-linear-factor-chi-2}) or obtained experimentally  (see Appendix~\ref{app: chi2-coeff}). These estimates are consistent with the approximation in  Eq.\,\eqref{eq:undepleting-pump-approx}. 
    
    \subsection{Below generation threshold}\label{sec:under-threshold-mode}
	
	Below the generation threshold, there is a natural limit to the minimum variance of the mode quadrature. In the undepleted-pump approximation, Eq.~\eqref{eq:zero_solution_Y2} limits the reduction of the squeezed-quadrature variance to a factor of two,  
	\begin{equation}
		\frac{0.25}{\langle {P^2}\rangle} = 1 + \eta^{(0)} < 2.
		\label{eq: 3 squeezing_limit}
	\end{equation}
	
\noindent
	This result corresponds to the well-known squeezing limit of 3 dB \cite{Milburn_OptComm_39_6_1981}: for squeezed quadrature $P$,
    corresponding to Eq.\,\eqref{eq:zero_solution_Y2}, $10\log_{10}(\corr{0.25}/\langle {P^2}\rangle) \leq10\log_{10}2\approx 3 $ dB.
	
	\noindent
	We now examine the validity of the undepleted-pump approximation in Eq.~\eqref{eq:undepleting-pump-approx}. If the pump-mode amplitude is well below its threshold value, then $\langle{X^2}\rangle - \langle {P^2}\rangle\sim 1$, and the condition \eqref{eq:undepleting-pump-approx} can be rewritten as $\lambda\ll 1$. Typical coefficient values of quadratic nonlinearity $g$ do not exceed $10^6$ \ Hz (see Appendix \ref{app: chi2-coeff}); the reverse lifetime of a photon, even in high-Q resonators, is at least $10^8$ Hz \cite{Lu_Optica_7_12_2020}. For these scales, the estimate is
	\begin{equation}\label{eq:lamda-estimation}
		\lambda = \frac{g^2}{\gamma^2} \approx 10^{-4}. 
	\end{equation}
	
	\noindent However, near the threshold, the signal mode, due to the growth of one of
the quadrature variances, begins to have a significant effect on the pump mode and the solution \eqref{eq:zero_solution_X2} -- \eqref{eq:zero_solution_pump} is no longer correct.

	  The estimate of $\lambda$ given above shows that $\lambda \ll 1$. Since $P$ is the squeezed quadrature, $\langle P^2\rangle$ remains of the order of the vacuum variance or smaller; therefore, the term $\lambda\langle P^2\rangle$ can be neglected in Eq.~\eqref{eq:dynamic_pump}.
      The approximate stationary solution for the variance
    of the antisqueezed quadrature, obtained from
    Eqs.~\eqref{eq:dynamic_X2}--\eqref{eq:dynamic_pump}, is
    
	\begin{equation}
		\langle {X^2}\rangle 
		\simeq 
		\frac{1}{4\lambda} 
		(\eta^{(0)} - 1 + \sqrt{(1 - \eta^{(0)})^2 + 2\lambda}).
		\label{eq:antisqueezing}
	\end{equation}
	
	\noindent
    Within the subthreshold regime, the function $\langle X^2\rangle(\eta^{(0)})$ in Eq.~\eqref{eq:antisqueezing} increases monotonically with $\eta^{(0)}$ and reaches its maximum as $\eta^{(0)}\to 1^-$. This gives the following estimate for the antisqueezed quadrature:

	\begin{equation}
		\langle {X^2}\rangle 
		\lesssim 
		\frac{1}{2\sqrt{2\lambda}}
		= 
		\frac{\gamma}{2\sqrt{2}g}.
		\label{eq:estimation_antisqueezing}
	\end{equation}
    If $\lambda \sim 10^{-4}$, this limit can be estimated as $\langle {X^2}\rangle \lesssim 10^2$.
	
\begin{figure*}[t]
    \centering
    \includegraphics[width=\textwidth]{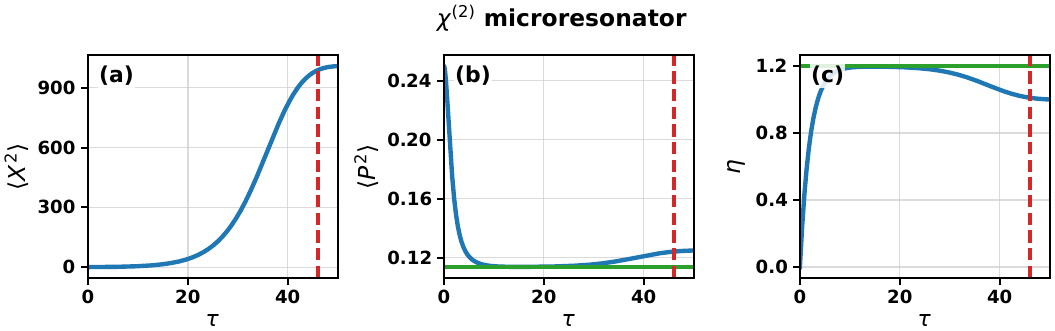}
    \caption{Transient dynamics in the $\chi^{(2)}$ microresonator for $\eta^{(0)}=1.2$ and $\lambda=10^{-4}$: \textbf{(a)} $\langle X^2\rangle$, \textbf{(b)} $\langle P^2\rangle$, and \textbf{(c)} the intracavity pump amplitude $\eta$, as functions of dimensionless time $\tau=\gamma t$. Blue curves show the numerical solution of Eqs.~\eqref{eq:dynamic_X2}--\eqref{eq:dynamic_pump}. Green horizontal lines in \textbf{(b,c)} show the fixed-pump values from Eqs.~\eqref{eq:zero_solution_Y2}--\eqref{eq:zero_solution_pump}. Red dashed vertical lines mark the estimate of $\tau_c$ from Eq.~\eqref{eq:coherent_time}. At $\tau=0$, $\langle X^2\rangle=\langle P^2\rangle=1/4$ and $\eta=0$. Constant external pumping is applied for $\tau\geq0$.}
    \label{fig:squeezing_dynamics_chi2}
\end{figure*}

	\subsection{Above generation threshold}\label{sec:above-threshold-mode}

In the above-threshold regime, the undepleted-pump approximation breaks down because the pump mode transfers a significant amount of energy to the signal mode. However, appreciable pump depletion does not occur immediately, but only after a characteristic time $\tau_c$, when the exponential growth of $\langle X^2\rangle$ described by Eq.~\eqref{eq:dynamic_X2} makes the criterion in Eq.~\eqref{eq:undepleting-pump-approx} no longer valid. This time can be estimated from Eq.~\eqref{eq:dynamic_X2} as
	\begin{equation}
		\tau_c 
		\approx
		\frac{1}{\eta^{(0)} - 1}
		\ln(\frac{1}{\lambda}).
		\label{eq:coherent_time}
	\end{equation}

	Consider the system of Eqs. \eqref{eq:dynamic_X2} -- \eqref{eq:dynamic_pump}, which takes into account the influence of the signal mode on the pumping mode. From the simplest analysis of this system, together with the above reasoning, it follows that for $\eta^{(0)} >1$ we have two characteristic evolution times: $\tau_0=1$ and $\tau_c$. At $\tau_0\ll\tau\ll \tau_c$, the condition in Eq.~\eqref{eq:undepleting-pump-approx} holds and the pump reaches an intermediate quasi-stationary state $\alpha_p = \alpha_p^{(0)} = \corr{\frac{2\varepsilon}{\gamma}}$.
At $\tau \gtrsim \tau_c$, the growth of the signal-mode quadrature variance causes appreciable pump depletion. In this regime, the validity of the classical-pump approximation requires additional assessment.
	
	In Fig.~\ref{fig:squeezing_dynamics_chi2}(a)--(c), a comparison of the numerical solution of the system \eqref{eq:dynamic_X2} -- \eqref{eq:dynamic_pump} and the qualitatively obtained results is presented. The numerical results show good agreement with the predicted characteristic timescales and the onset of pump depletion.

	During the time interval $\tau_0 < \tau <\tau_c$, the value of $\langle {X^2}\rangle$ experiences exponential growth, and the squeezed-quadrature variance, which equals $\langle P^2\rangle$ because $\langle P\rangle=0$, approaches the intermediate value
	
	\begin{equation}
		 \langle (\Delta P)^2 \rangle = \frac{0.25}{1 + \eta^{(0)}}.
		\label{eq:oversqueezing}
	\end{equation}
\noindent
   Although this value is no longer constant for $\langle{P^2}\rangle$ above the threshold, it can be obtained at an intermediate stage of evolution; also, it can be seen from Eq. \eqref{eq:oversqueezing} that the squeezing limit of 3 dB can be overcome during this interval.

	 For $\eta^{(0)}=2.0$ and $\lambda=10^{-4}$, the lifetime $\tau_c$ of the transient squeezed state is estimated as $\tau_c\sim10$, which, at $\gamma=10^8$ Hz, corresponds to $t_c=\tau_c/\gamma\sim100$ ns. The time will decrease as \(\eta^{(0)}\) exceeds the threshold.
		

\section{Generation of squeezed states via Kerr interaction}\label{sec:qubic_nonlinear}
	
	In this section, we study the dynamics of squeezed state generation in a Kerr microresonator. We consider a system of three modes: the signal mode with frequency $\omega$ and two pump modes with frequencies $\omega_1 = \omega -\delta \omega$ and $\omega_2 =\omega+\delta \omega$. Pump mode radiation is injected into a resonator by an external waveguide. The central mode arises as a result of degenerate dual-pump spontaneous four-wave mixing (DP-SFWM) in a cubic nonlinear medium.
	
	The Lindblad equation, Hamiltonian, and dissipator for the three-mode system take the following forms:
	
	\begin{equation}
		\frac{d\hat{\rho}_T}{dt} 
		= 
		-i[\hat{H}_T, \hat{\rho_T}] + L(\hat{a}) + L(\hat{a}_1) + L(\hat{a}_2),
		\label{eq:lindblad_equasion_chi3}
	\end{equation}

    \begin{equation}
\begin{aligned}
\hat{H}_T ={}&
\frac{g}{2}\left(
\hat{a}^{\dagger 2}\hat{a}_1\hat{a}_2
+\hat{a}^{2}\hat{a}_1^\dagger\hat{a}_2^\dagger
\right)\\
&-\varepsilon_1\left(
e^{-i\pi/4}\hat{a}_1^\dagger
+e^{i\pi/4}\hat{a}_1
\right)\\
&-\varepsilon_2\left(
e^{-i\pi/4}\hat{a}_2^\dagger
+e^{i\pi/4}\hat{a}_2
\right),
\end{aligned}
\label{eq:hamiltonian_chi3}
\end{equation}
     
	\begin{equation}
		L(\hat{a}) = \frac{\gamma}{2}([\hat{a}\hat{\rho_T}, \hat{a}^{\dagger}] + [\hat{a}, \hat{\rho_T} \hat{a}^{\dagger}]).
		\label{eq:collaps_op_chi3}
	\end{equation}

    \noindent
\begin{figure*}[t]
    \centering
    \includegraphics[width=\textwidth]{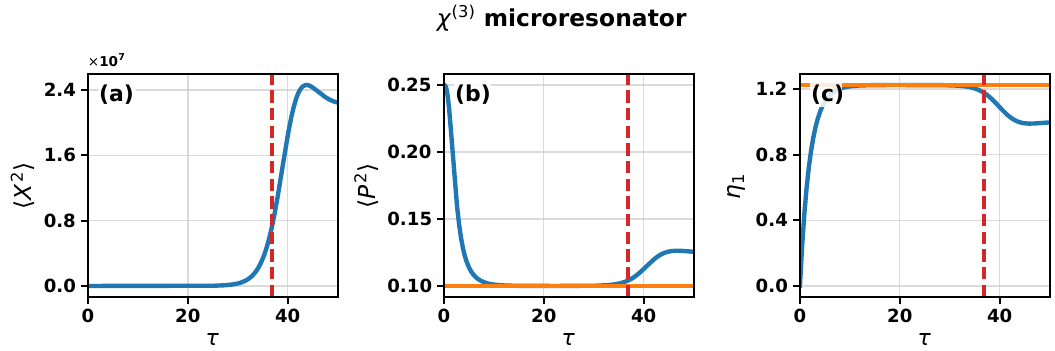}
    \caption{Transient dynamics in the three-mode $\chi^{(3)}$ microresonator for $\eta_1^{(0)}=\eta_2^{(0)}=\sqrt{1.5}$ and $\gamma/g=10^8$: \textbf{(a)} $\langle X^2\rangle$, \textbf{(b)} $\langle P^2\rangle$, and \textbf{(c)} the intracavity pump amplitude $\eta_1$, as functions of dimensionless time $\tau=\gamma t$. Blue curves show the numerical solution of Eqs.~\eqref{eq:dynamic_final_X2_chi3}--\eqref{eq:dynamic_final_blue_pump_chi3}. Orange horizontal lines in \textbf{(b,c)} show the fixed-pump values from Eqs.~\eqref{eq:solution_Y2_chi3}--\eqref{eq:solution_pumps_chi3}. Red dashed vertical lines mark the estimate of $\tau_c$ from Eq.~\eqref{eq:coherent_evolution_chi3}. At $\tau=0$, $\langle X^2\rangle=\langle P^2\rangle=1/4$ and $\eta_1=\eta_2=0$. Constant external pumping is applied for $\tau\geq0$.}
    \label{fig:squeezing_dynamics_chi3}
\end{figure*}

	The external drive amplitudes $\varepsilon_i$ are given by
	
	\begin{equation}
		\varepsilon_i = \sqrt{\frac{P_i\gamma_c}{\hbar \omega_i}}\,,
		\label{eq:eps_expression_chi3}
	\end{equation}

     \noindent
     where $\omega_i$ is the frequency of the pump $i$, $P_i$ is its power in the waveguide,  $g$ is the cubic nonlinearity coefficient, $\gamma_c$ is the coupling decay rate, and $\gamma$ is the photon decay rate in the resonator. For simplicity, we assume that the photon lifetime in the resonator is approximately the same for all three frequencies, since $\gamma$ does not change significantly for a frequency shift of $\delta\omega$.

As in the quadratic case, we treat the pump modes as coherent and replace their operators by classical amplitudes,
	
	\begin{equation}
		\hat{a}_i \xrightarrow{} e^{i\pi/4}\alpha_i \ \ \ (i = 1, 2),
		\label{eq:undepleted_pump_approx_chi3}
	\end{equation}
	where $\alpha_1, \alpha_2$ are real.
\noindent	
	 In the Heisenberg picture, we obtain a system of differential equations
for the expectation values of the squared signal-mode field operators
and for the two pump amplitudes:
	
	\begin{equation}
		\frac{d\langle {X^2}\rangle}{dt} 
		= 
		(2\alpha_1\alpha_2 g - \gamma)\langle {X^2}\rangle + \frac{\gamma}{4},
		\label{eq:dynamic_X2chi3}
	\end{equation}
	
	\begin{equation}
		\frac{d\langle {P^2}\rangle}{dt} 
		= 
		-(2\alpha_1\alpha_2 g + \gamma)\langle {P^2}\rangle 
		+ 
		\frac{\gamma}{4},
		\label{eq:dynamic_Y2chi3}
	\end{equation}
	
	\begin{equation}
		\frac{d\alpha_{1,2}}{dt} 
		= 
		\varepsilon_{1,2} 
		- 
		\frac{1}{2} g\alpha_{2,1}(\langle {X^2}\rangle 
		- 
		\langle {P^2}\rangle) - \frac{1}{2} \gamma \alpha_{1,2}.
		\label{eq:dynamic_pumps_chi3}
	\end{equation}
	\noindent
	Introducing the dimensionless time $\tau =\gamma t$ and normalized pump amplitudes $\eta_i=\sqrt{\frac{2 g}{\gamma}}\alpha_i$ allows us to rewrite the system of equations as follows:
	
	\begin{equation}
		\frac{d\langle {X^2}\rangle}{d\tau} 
		= 
		(\eta_1\eta_2 - 1)\langle {X^2}\rangle + \frac{1}{4},
		\label{eq:dynamic_final_X2_chi3}
	\end{equation}
	
	\begin{equation}
		\frac{d\langle {P^2}\rangle}{d\tau} 
		= 
		-(\eta_1\eta_2 + 1)\langle {P^2}\rangle + \frac{1}{4},
		\label{eq:dynamic_final_Y2_chi3}
	\end{equation}
	
	\begin{equation}
		\frac{d\eta_1}{d\tau} 
		=
		\frac{1}{2}\eta_1^{(0)} 
		- 
		\frac{g}{\gamma}\frac{\eta_2}{2}(\langle {X^2}\rangle 
		- 
		\langle {P^2}\rangle) - \frac{1}{2}\eta_1,
		\label{eq:dynamic_final_red_pump_chi3}
	\end{equation}
	\begin{equation}
		\frac{d\eta_2}{d\tau} 
		= 
		\frac{1}{2}\eta_2^{(0)} 
		- 
		\frac{g}{\gamma}\frac{\eta_1}{2}
		(\langle {X^2}\rangle -	\langle {P^2}\rangle) 
		-
		\frac{1}{2}\eta_2.
		\label{eq:dynamic_final_blue_pump_chi3}
	\end{equation}
	\noindent
	Here $\eta_i^{(0)} =\sqrt{\frac{8g}{\gamma}}\frac{\varepsilon_i}{\gamma} =\sqrt{\frac{2g}{\gamma}}\alpha_i^{(0)}$, where $\alpha_i^{(0)}$ is the stationary value of the $i-$th pump amplitude in an optically linear medium.
	\noindent
	The generation threshold is reached when
	\begin{equation}
		\eta_1^{(0)} \eta_2^{(0)} = 1.
		\label{eq:threshold_chi33}
	\end{equation}
	
	In real systems, for example in high-Q resonators in the optical band, the condition $\gamma\gg g$ is fulfilled. Well below the generation threshold, the stationary quadrature variances and pump amplitudes are
	\begin{equation}
		\langle {X^2}\rangle = \frac{\corr{0.25}}{1-\eta_1^{(0)}\eta_2^{(0)}},
		\label{eq:solution_X2_chi3}
	\end{equation}
	
	\begin{equation}
		\langle {P^2}\rangle = \frac{\corr{0.25}}{\eta_1^{(0)}\eta_2^{(0)} + 1},
		\label{eq:solution_Y2_chi3}
	\end{equation}
	
	\begin{equation}
		\eta_i = \eta_i^{(0)}.
		\label{eq:solution_pumps_chi3}
	\end{equation}
	\noindent
	Figure~\ref{fig:squeezing_dynamics_chi3}(b) shows the transient quasi-stationary variance in the above-threshold regime ($\eta_1\eta_2 > 1$). As in the $\chi^{(2)}$ case, the variance governed by Eq.~\eqref{eq:dynamic_final_Y2_chi3} is quasi-stationary over the time interval inferred from Eq.~\eqref{eq:dynamic_final_X2_chi3},
	\begin{equation}
		1 < \tau < \tau_c \approx \frac{1}{\eta_1^{(0)}\eta_2^{(0)} - 	1}\ln(\frac{\gamma}{g}).
		\label{eq:coherent_evolution_chi3}
	\end{equation}

    \noindent
	For typical parameter values
	$\gamma\sim 10^8 \ \text{Hz}, g\sim 1 \ \text{Hz}$ (see Appendix \ref{app: chi3-coeff}, \ \cite{Ramelow_RPL_122_153906_2019})
at $\eta_1\eta_2 = 1.5$, the minimum variance is smaller than the subthreshold limit by a factor of 1.25. The characteristic time is estimated as  \(\tau_c\approx 36\), which corresponds to \(t_c\approx 360\) ns.

	\subsection{Effect of parasitic single-pump four-wave mixing}
	The model, taking into account only three modes in the resonator, is a significant approximation; it does not consider all possible processes of four-wave mixing. It is reasonable to add 2 more modes to our system at frequencies $\omega_m = \omega - 2\delta \omega$ and $\omega_n= \omega+ 2\delta \omega$. The energy conservation conditions for these single-pump four-wave-mixing processes are
$$2\omega_1 =\omega_m + \omega, \quad 2\omega_2=\omega + \omega_n.$$

\noindent
The parasitic process is single-pump spontaneous four-wave mixing (SP-SFWM). The Hamiltonian includes both dual-pump and single-pump contributions,
	\begin{equation}
		\hat{H} = \hat{H}_{DP-SFWM} + \hat{H}_{SP-SFWM},
		\label{eq:hamiltonian_with_parasitic_processes}
	\end{equation}
	
	\begin{equation}
		\hat{H}_{SP-SFWM} 
		= i\frac{g\alpha_1^2}{2}(\hat{a}^{\dagger}\hat{a}_m^{\dagger} - \hat{a}\hat{a}_m) + 
		i\frac{g\alpha_2^2}{2}(\hat{a}^{\dagger}\hat{a}_n^{\dagger} - \hat{a}\hat{a}_n).
		\label{eq:hamiltonian_SP_SFWM}
	\end{equation}
\noindent
We take into account one parasitic channel in the five-mode system
(see Appendix~\ref{app:parasitic-modes} for the Hamiltonian used in this reduced model). This reduced model captures the
leading detrimental effect on squeezing, namely the generation of additional
photons in the signal modes that are not correlated with the photons produced
by the target four-wave-mixing process~\cite{Seifoory_PRA_105_033524_2022}.
Following Ref.~\cite{Seifoory_PRA_105_033524_2022}, only this pair of
parasitic modes is retained. Appropriate group-velocity-dispersion
engineering may suppress higher-order parasitic channels by making them
strongly phase-mismatched~\cite{Fujii_OptLett_44_3146_2019}, thus motivating the five-mode approximation employed here. Selective resonance splitting has also been demonstrated to suppress unwanted nonlinear processes in nanophotonic microresonators~\cite{Ulanov_NatCommun_16_10791_2025}.

    Still considering the pumps to be classical, we can write a linear system of differential equations for the average values of operators of the form $\hat{a}_i\hat{a}_j$ and
$\hat{a}_i\hat{a}_j^\dagger$ (see Appendix \ref{app:parasitic-modes}). The stationary solution of such a system allows for the calculation of the values of signal mode variances. In this case, the generation threshold is determined by the onset of instability, when the largest real part of the eigenvalues of the dynamical matrix reaches zero.
	
\begin{figure}[H]
		\centering
		\includegraphics[width=0.99\linewidth]{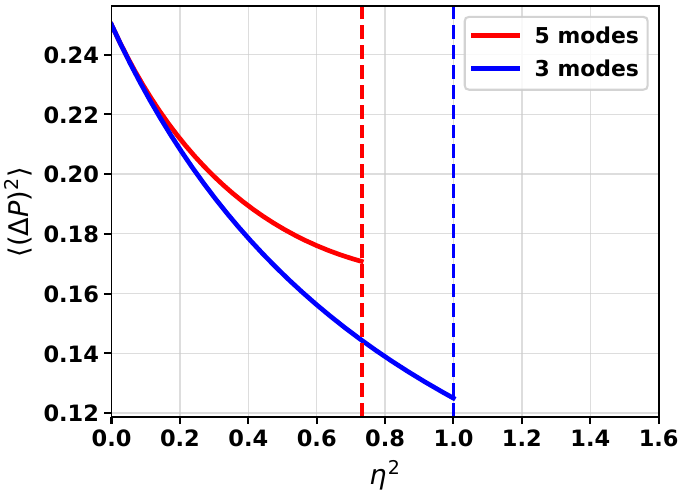}
		\caption{The dependence of the variance of the squeezed quadrature of the signal mode $\langle{(\Delta P)^2}\rangle$ on the parameter $\eta^2$. The blue curve visualizes Eq. \eqref{eq:solution_Y2_chi3} for the three-mode model. The red curve shows the result for the five-mode model described in Appendix~\ref{app:parasitic-modes}. Vertical lines mark the threshold values of $\eta^2$ for the two models.}
        \label{fig:parasitic-mode-chi-3}
	\end{figure}

	Figure \ref{fig:parasitic-mode-chi-3} shows a comparison of the calculation of the signal mode variances in a three-mode model and in a model for a system of five modes, which takes into account the process described above involving lateral parasitic modes, for the case of pumps of the same power ($\eta_1 = \eta_2 = \eta)$.

From this comparison, we draw the following qualitative
conclusions. First, taking SP-SFWM into account increases
the variance of the squeezed quadrature and therefore
reduces the squeezing along this axis. The variance of
the orthogonal quadrature also increases. Second, within the reduced model considered here, the coupled quantum-mode subsystem becomes unstable at lower pump amplitudes.
For the symmetric case considered here,
$\eta_1=\eta_2=\eta$, the threshold in the five-mode
model is $\eta^2=\sqrt{3}-1\approx 0.732$, whereas
the three-mode model gives $\eta^2=1.0$.

	Thus, we demonstrate the calculation of stationary second-order moments and signal-mode quadrature variances within the five-mode model. At the same time, the linearization of the equations turned out to be possible only under the condition of constant pumping amplitudes, which, however, is acceptable when operating significantly below the generation threshold. The calculation took into account only one parasitic process of four-wave mixing, which negatively affects the squeezing ratio. The inclusion of this parasitic process lowers the generation threshold and reduces the achievable squeezing of the signal mode.

\section{Conclusion}\label{sec:discussion}

 The proposed model describes the dynamics of intracavity squeezed-state generation in nonlinear microresonators, taking into account the back-action of the signal mode on the classical pump amplitudes. Below the generation threshold, as well as at the initial stage immediately above the threshold, the model yields results consistent with known limits (e.g., the 3~dB squeezing limit). Moreover, under above-threshold pumping, the model describes transient intracavity squeezing beyond 3 dB, followed by its reduction as pump depletion becomes appreciable.

Far below the threshold, the undepleted pump approximation remains valid, simplifying the analysis. However, near and above the generation threshold, pump depletion can become important for the subsequent evolution of the system. Our approach provides a way to estimate the time scale on which pump depletion becomes significant, thereby delineating the regime of validity for the undepleted pump assumption.

While our current model neglects certain effects (such as material dispersion, cross- and self-phase modulation in the Kerr medium, and additional parasitic mode interactions beyond one sideband pair), it can be extended to incorporate these factors. Including dispersion and additional parasitic processes could improve estimates of the achievable intracavity squeezing in integrated microresonators and help identify strategies to suppress unwanted nonlinear interactions, for example by tailoring the resonator mode spectrum.

In summary, we have shown that by accounting for the dynamic interaction between the signal and pump modes in an optical resonator, one can explore regimes of squeezed light generation that go beyond the traditional undepleted pump approximation. Our findings show that the conventional 3 dB intracavity squeezing limit can be exceeded transiently under above-threshold pumping, while pump depletion limits the duration of this regime. These results may help identify operating regimes and timescales for exploiting transient intracavity squeezing in microresonator-based devices. Further work will involve extending the model to incorporate additional physical effects and exploring control techniques to sustain and utilize the enhanced squeezing in practical applications.

\begin{acknowledgments}
This work was supported by the Russian Science Foundation (grant No.~26-72-10176).
\end{acknowledgments}

\section*{Data availability}
The data supporting the findings of this study are available from the corresponding authors upon reasonable request.

\section*{Conflict of interest}
The authors declare no conflict of interest.

\appendix
    \section{Estimation of quadratic nonlinear coefficient}
    \label{app: chi2-coeff}
    For the $\chi^{(2)}$ interaction, we estimate the nonlinear coefficient \(g_2\) of the Hamiltonian using the expression from Ref.~\cite{Okoth_PRA_99_4_2019},
    \begin{equation}\label{eq:non-linear-factor-chi-2}
        g_2 \sim \frac{1}{4\sqrt{2}n_s^2}\frac{\sqrt{\hbar \omega_p}\chi^{(2)} \omega_s}{\sqrt{V \varepsilon_0}n_p^2},  
    \end{equation}
    where $\omega_s$ is the signal-mode angular frequency, $\omega_p$ is the pump-mode angular frequency, $\chi^{(2)}$ is the second-order nonlinear susceptibility, $\hbar$ is the reduced Planck constant, $V$ is the effective nonlinear interaction volume, $\varepsilon_0$ is the vacuum permittivity, and $n_s$ and $n_p$ are the refractive indices at the signal and pump frequencies, respectively.
    For representative parameters of $\mathrm{LiNbO_3}$ microresonators, $\omega_s \sim 10^{15}\,\mathrm{s}^{-1}$, $V \approx 5\cdot 10^{-15}\,\mathrm{m}^3$~\cite{gao2021broadband}, $n_s \sim 2.2$ and $n_p = 2.26$~\cite{zelmon1997infrared}, and $\chi^{(2)} = 2\cdot10^{-11}\,\mathrm{m/V}$~\cite{schiek2012absolute}, we obtain $g_2 \sim 10^6\,\mathrm{Hz}$.

    \section{Estimation of the Kerr nonlinear coefficient}
    \label{app: chi3-coeff}
    For the $\chi^{(3)}$ interaction, the analogous estimate for the nonlinear coefficient $g_3$ is~\cite{Ramelow_RPL_122_153906_2019}
    \begin{equation}\label{eq:non-linear-factor-chi-3}
        g_3 \sim \frac{\hbar \omega^2\chi^{(3)}}{{V \varepsilon_0 n^4}},
    \end{equation}
    where $\omega$ is the signal-mode angular frequency, $\chi^{(3)}$ is the third-order nonlinear susceptibility, $\hbar$ is the reduced Planck constant, $V$ is the effective nonlinear interaction volume, $\varepsilon_0$ is the vacuum permittivity, and $n$ is the refractive index.
  For representative parameters of Si$_3$N$_4$ microresonators, $\omega \sim 10^{15}\,\mathrm{s}^{-1}$, $V \sim 10^{-15}\,\mathrm{m}^3$~\cite{Seifoory_PRA_105_033524_2022}, $n = 2.0$~\cite{luke2015broadband}, and $\chi^{(3)} \sim 3.4\cdot10^{-21}\,\mathrm{m^2/V^2}$~\cite{Ikeda_OptExpress_16_12987_2008}, we obtain $g_3 \sim 1\,\mathrm{Hz}$, consistent with the estimate in Ref.~\cite{Seifoory_PRA_105_033524_2022}.

\section{Evolution of a system with parasitic modes}
	\label{app:parasitic-modes}

    \noindent
	This section presents a calculation of the evolution of a five-mode system, taking into account DP-SFWM and SP-SFWM. We assume zero detunings and a common decay rate $\gamma$ for the signal mode and both parasitic modes $m$ and $n$.
	
	With fixed classical pump amplitudes, the density matrix describes the signal and the two parasitic modes $m$ and $n$. The Lindblad equation, Hamiltonian, and dissipator take the following forms:
	\begin{equation}
		\frac{d\hat{\rho}_T}{dt} = -i[\hat{H}_T, \hat{\rho}_T] + L(\hat{a}) + L(\hat{a}_m) + L(\hat{a}_n),
	\end{equation} 
	
	\begin{equation}
\begin{aligned}
\hat{H}_T ={}& \frac{ig}{2}\alpha_1\alpha_2\big((\hat{a}^{\dagger})^2 - \hat{a}^2\big) \\
&+ \frac{ig}{2}\alpha_1^2(\hat{a}^{\dagger}\hat{a}_m^{\dagger} - \hat{a}\hat{a}_m) \\
&+ \frac{ig}{2}\alpha_2^2(\hat{a}^{\dagger}\hat{a}_n^{\dagger} - \hat{a}\hat{a}_n),
\end{aligned}
	\end{equation}
	
	\begin{equation}
		L(\hat{a}) = \frac{\gamma}{2}([\hat{a}\hat{\rho_T}, \hat{a}^{\dagger}] + [\hat{a}, \hat{\rho_T} \hat{a}^{\dagger}]).
	\end{equation}

    \noindent
	In the Heisenberg picture, the Lindblad master
equation yields the following equation of motion
for an arbitrary system operator $\hat{A}$:
	\begin{equation}
		\frac{d\hat{A}}{dt} = i[\hat{H}_T, \hat{A}] + \frac{\gamma}{2}\sum_{\hat b\in\{\hat a,\hat a_m,\hat a_n\}}(\hat b^{\dagger}[\hat A,\hat b]+[\hat b^{\dagger},\hat A]\hat b).
	\end{equation}

    \noindent
	We substitute products of the form $\hat{a}_i\hat{a}_j$ and $\hat{a}_i\hat{a}_j^{\dagger}$ for $\hat{A}$ and use the dimensionless variables $\tau=\gamma t$ and $\eta_i^2=2g\alpha_i^2/\gamma$ to obtain the following system:
	\begin{equation}
		\frac{d\hat{a}^2}{d\tau} = - \hat{a}^2 + \eta_1\eta_2\hat{a}\hat{a}^{\dagger} 
		+ \frac{\eta_1^2}{2}\hat{a}\hat{a}_m^{\dagger} + \frac{\eta_2^2}{2}\hat{a}\hat{a}_n^{\dagger}
		- \frac{\eta_1\eta_2}{2},
		\label{eq:parasitic_first_moment_equation}
	\end{equation}
	\begin{equation}
		\begin{aligned}
			\frac{d(\hat{a}\hat{a}^{\dagger})}{\corr{d\tau}} = - \hat{a}\hat{a}^{\dagger} + \frac{\eta_1^2}{4}(\hat{a}\hat{a}_m + 
			\hat{a}^{\dagger}\hat{a}_m^{\dagger}) \\+ \frac{\eta_2^2}{4}(\hat{a}\hat{a}_n + 
			\hat{a}^{\dagger}\hat{a}_n^{\dagger}) + \frac{\eta_1\eta_2}{2}((\hat{a}^{\dagger})^2 + 
			\hat{a}^2) + 1,
		\end{aligned}
	\end{equation}
	
	\begin{equation}
		\frac{d(\hat{a}\hat{a}_m)}{d\tau} = - \hat{a}\hat{a}_m + 
		\frac{\eta_1^2}{4}(\hat{a}^{\dagger}\hat{a} + \hat{a}_m^{\dagger}\hat{a}_m + 1) + 
		\frac{\eta_2^2}{4}\hat{a}_m\hat{a}_n^{\dagger} + 
		\frac{\eta_1\eta_2}{2}\hat{a}^{\dagger}\hat{a}_m,
	\end{equation}
	\begin{equation}
		\frac{d(\hat{a}\hat{a}_m^{\dagger})}{d\tau} = - \hat{a}\hat{a}_m^{\dagger} + 
		\frac{\eta_1^2}{4}(\hat{a}^2 + (\hat{a}_m^{\dagger})^2) + 
		\frac{\eta_2^2}{4}\hat{a}_m^{\dagger}\hat{a}_n^{\dagger} + 
		\frac{\eta_1\eta_2}{2}\hat{a}^{\dagger}\hat{a}_m^{\dagger},
	\end{equation}
	\begin{equation}
		\frac{d(\hat{a}\hat{a}_n)}{d\tau} = - \hat{a}\hat{a}_n + 
		\frac{\eta_2^2}{4}(\hat{a}^{\dagger}\hat{a} + \hat{a}_n^{\dagger}\hat{a}_n + 1) + 
		\frac{\eta_1^2}{4}\hat{a}_n\hat{a}_m^{\dagger} + 
		\frac{\eta_1\eta_2}{2}\hat{a}^{\dagger}\hat{a}_n,
	\end{equation}
	
	\begin{equation}
		\frac{d(\hat{a}\hat{a}_n^{\dagger})}{d\tau} = - \hat{a}\hat{a}_n^{\dagger} + 
		\frac{\eta_2^2}{4}(\hat{a}^2 + (\hat{a}_n^{\dagger})^2) + 
		\frac{\eta_1^2}{4}\hat{a}_n^{\dagger}\hat{a}_m^{\dagger} + 
		\frac{\eta_1\eta_2}{2}\hat{a}^{\dagger}\hat{a}_n^{\dagger},
	\end{equation}
	
	\begin{equation}
		\frac{d\hat{a}_m^2}{d\tau} = -\hat{a}_m^2 + \frac{\eta_1^2}{2}\hat{a}^{\dagger}\hat{a}_m,
	\end{equation}
	
	\begin{equation}
		\frac{d(\hat{a}_m\hat{a}_m^{\dagger})}{d\tau} = -\hat{a}_m\hat{a}_m^{\dagger} +
		\frac{\eta_1^2}{4}(\hat{a}^{\dagger}\hat{a}_m^{\dagger} + \hat{a}\hat{a}_m) + 1,
	\end{equation}
	
	\begin{equation}
		\frac{d(\hat{a}_m\hat{a}_n)}{d\tau} = -\hat{a}_m\hat{a}_n + 
		\frac{\eta_1^2}{4}\hat{a}^{\dagger}\hat{a}_n + \frac{\eta_2^2}{4}\hat{a}^{\dagger}\hat{a}_m,
	\end{equation}
	
	\begin{equation}
		\frac{d(\hat{a}_m\hat{a}_n^{\dagger})}{d\tau} = - \hat{a}_m\hat{a}_n^{\dagger} + 
		\frac{\eta_1^2}{4}\hat{a}^{\dagger}\hat{a}_n^{\dagger} +
		\frac{\eta_2^2}{4}\hat{a}\hat{a}_m,
	\end{equation}
	
	\begin{equation}
		\frac{d\hat{a}_n^2}{d\tau} = - \hat{a}_n^2 + \frac{\eta_2^2}{2}\hat{a}^{\dagger}\hat{a}_n,
	\end{equation}
	
	\begin{equation}
		\frac{d(\hat{a}_n\hat{a}_n^{\dagger})}{d\tau} = -\hat{a}_n\hat{a}_n^{\dagger}
		+ \frac{\eta_2^2}{4}(\hat{a}^{\dagger}\hat{a}_n^{\dagger} + \hat{a}\hat{a}_n) + 1.
		\label{eq:parasitic_last_moment_equation}
	\end{equation}
	\noindent
	 Equations~\eqref{eq:parasitic_first_moment_equation}--\eqref{eq:parasitic_last_moment_equation} describe the Heisenberg-picture evolution of second-order operator products. Taking expectation values yields a closed linear system for the corresponding moments. We write this system in matrix form to obtain its stationary solution. In the stable below-threshold regime, the mean amplitudes of the signal and parasitic modes vanish. Consequently, the quadrature variances coincide with their second moments, $\langle(\Delta X)^2\rangle=\langle X^2\rangle$ and $\langle(\Delta P)^2\rangle=\langle P^2\rangle$, which can be calculated using the following operator identities:
    $$
    \hat{X}^2 = \frac{1}{4}(\hat{a}^{2} + (\hat{a}^{\dagger})^2 + \hat{a}\hat{a}^{\dagger} + \hat{a}^{\dagger}\hat{a}),
    $$
    $$
    \hat{P}^2 = \frac{1}{4}(-\hat{a}^{2} - (\hat{a}^{\dagger})^2 + \hat{a}\hat{a}^{\dagger} + \hat{a}^{\dagger}\hat{a}).
    $$

	 \bibliography{references}
\end{document}